\documentclass[reprint]{revtex4-1}

\usepackage{graphicx}
\usepackage{amsmath}
\usepackage{dcolumn}
\usepackage{bm}
\usepackage[colorlinks=true,linkcolor=blue,citecolor=blue,urlcolor=blue]{hyperref}%

\begin{document}


\title{Finite-element simulations of hysteretic ac losses\\
in a magnetically coated superconducting tubular wire\\
subject to an oscillating transverse magnetic field}

\author{Y. A.~Genenko}
\author{H. Rauh}%
\email[Electronic mail:\,]{hera@tgm.tu-darmstadt.de}
\author{S. Kurdi}
\affiliation{%
Institut f\"ur Materialwissenschaft, Technische Universit\"at Darmstadt, Jovanka-Bontschits-Strasse 2,\\ 
D-64287 Darmstadt, Germany
}%

\date{\today}

\begin{abstract}
Numerical simulations of hysteretic ac losses in a tubular superconductor/paramagnet 
heterostructure subject to an oscillating transverse magnetic field are performed within the quasistatic 
approach, calling upon the COMSOL finite-element software package and exploiting magnetostatic-electrostatic 
analogues. It is shown that one-sided magnetic shielding of a thin, type-II superconducting tube by a coaxial
paramagnetic support results in a slight increase of hysteretic ac losses as compared to those for a vacuum 
environment, when the support is placed inside; a spectacular shielding effect with a possible reduction of 
hysteretic ac losses by orders of magnitude, however, ensues,  depending on the magnetic permeability and 
the amplitude of the applied magnetic field, when the support is placed outside.  
\end{abstract}

\pacs{74.25.N-, 74.25.Op, 74.78.Fk, 85.25.-j}
\maketitle

\section{INTRODUCTION}
\label{intro} 
  
Cylindrical heterostructures made up of superconductor and paramagnet constituents find
use in various technological applications, power transmission cables involving second-generation
high-temperature superconductor multistrand wires and electromagnetic coils based on coated 
conductors\cite{Malozemoff2010} or bulk MgB$_2$/Fe filaments\cite{Sumption2002} playing a prominent role 
thereby. The combined magnetic shielding properties of these kinds of constituents render beneficial 
effects like a reduction of hysteretic ac losses\cite{Sumption2002,MajorosPhysC-2} or an enhancement 
of the (field-dependent) critical current,\cite{Horvat2002,Genenko2004,Genenko2005,Yampolskii2005}
to name just a few. 
Given specific material characteristics and geometries, coaxial superconductor/paramagnet
heterostructures may even disclose the exceptional hallmarks of magnetic 
cloaks.\cite{NJP13,Science2012,Narayana2012,Yampolskii2014-1,Yampolskii2014-2}

Hysteretic ac losses in unshielded superconductors of cylindrical shapes have been established for 
the cases of an oscillating transport current imposed or an oscillating transverse magnetic field 
applied\cite{Ashkin1970,Majoros2014} as well as for scenarios with both types of excitations in operation
at the same time.\cite{Ruiz2012,Ruiz2013} By comparison, hysteretic ac losses in superconductor strips on 
magnetic supports were until recently discussed controversially, some experiments demonstrating a monotonic 
rise of hysteretic ac losses with increasing permeability of soft-magnetic constituents,\cite{Claassen2008}
others showing a contingent reduction of such losses due to soft-magnetic supports.\cite{Duck2003,Grilli2007}
In fact, hysteretic ac losses may either be depressed or enhanced depending on the geometrical and material
characteristics of the magnetic environment\cite{Genenko2000,AladSUST2009,Genenko2010} and the amplitude
of the applied magnetic field.\cite{Mawatari2008,GomorySUST2010}
Taking advantage of
magnetostatic-electrostatic analogues,\cite{Genenko2009} satisfactory agreement between numerically 
simulated\cite{Genenko2011} and experimentally determined\cite{Suenaga2008} hysteretic ac
losses in a planar bilayer superconductor/paramagnet heterostructure subject to an oscillating
transverse magnetic field could be achieved, with proper dependences on the permeability of the 
magnetic support $\mu $ and the amplitude  of the applied magnetic field $H_a$. Thorough accounts of 
both analytical and numerical methods for modelling electromagnetic properties of high-temperature 
superconductors, including hysteretic ac losses in the presence of paramagnetic supports, offer two 
up-to-date reviews.\cite{Mikitik2013, Grilli2014}

Regarding tubular superconductor/paramagnet heterostructures forged by polygonal assemblies of coated 
superconductor strips, great effort has been spent on finite-element simulations of hysteretic ac losses 
for an oscillating transport current imposed.\cite{Amemiya2007,Miyagi2007,Amemiya2010} Accordingly, when 
assuming a nonlinear current-voltage characteristic of the superconductors and a constant permeability
$\mu$ of the magnetic constituents, the transport current hysteretic ac loss with the magnetic supports 
placed inside turns out to be far less than with these supports placed outside.\cite{Miyagi2007} 
Increasing $\mu$ in configurations with inner magnetic supports leads to enhancements of the hysteretic 
ac loss;\cite{Amemiya2007} an effect which saturates for a relative magnetic permeability near 
$\mu/\mu_0 \simeq 100$, as anticipated from previous analysis.\cite{Genenko2000,GenenkoJAP2002} 
Incorporation of a nonlinear, but reversible, 
field-dependence of the magnetic permeability $\mu$ has merely little bearing on the saturated magnetic 
property, as comparative modellings using fixed values of $\mu$ show,\cite{Amemiya2007} very much like for a
single magnetically shielded superconductor strip.\cite{Ma2013} Advanced studies of a power transmission 
cable involving two coaxial sets of coated superconductor strips -- with hysteretic magnetization losses 
suffered by the magnetic constituents allowed for too -- unfold that these losses control the dissipation 
of electromagnetic energy of the entire heterostructure.\cite{Amemiya2010}

Although multistrand wires and electromagnetic coils are obviously exposed to transverse
magnetic fields, only few works seem to have addressed the electromagnetic behaviour of
tubular superconductor/paramagnet heterostructures in oscillating applied magnetic fields. An
elaborate numerical study of the shielding properties of such heterostructures, for example,
assumes a nonlinear current-voltage characteristic of the superconductors and a reversible
field-dependence of the permeability of the magnetic constituents $\mu$, apart from including the
field-effect on the critical current density $j_c$.\cite{Lousberg2010} Hence, configurations 
with the paramagnet layers placed outside the superconductor constituents exhibit much stronger 
shieldings than those with the respective layers placed inside. Based on an elegant analytical 
approach,\cite{Mawatari2008,Mawatari2009} recent work probes the effect of outer magnetic 
supports on the hysteretic ac loss in a power transmission cable assembled from curved 
superconductor/soft-magnet tapes conforming to a cylindrical shape.\cite{He2014} Both types of 
constituents are deemed infinitesimally thin, the magnetic supports being characterized 
by an infinite permeability, $\mu\rightarrow\infty$. A realistic enhancement of the hysteretic 
ac loss due to the magnetic supports obtains for a transport current imposed, consistent with 
investigations before;\cite{Amemiya2007,Miyagi2007} results derived for an applied magnetic 
field of radial symmetry, however, scarcely hit considering the fact that an external magnetic 
field must be source free. Significant progress exemplifies a theoretical analysis of the 
electromagnetic response of a cylindrical tubular wire represented by an infinitesimally  thin 
superconductor constituent subject to an oscillating transverse magnetic field.\cite{Mawatari2011} 
In a description where the superconductor is delineated by the sheet current $J$, with a 
(field-independent) critical value $J_c$, the profiles of the magnetic field, the field of
first penetration of magnetic flux and the hysteretic ac losses that ensue pave the way towards
research on the electromagnetic behaviour of tubular heterostructures embracing superconductor 
as well as paramagnet constituents. We here extend this ansatz for a coated superconducting tubular 
wire, with a coaxial paramagnetic support, by making recourse to Bean's model of the critical 
state\cite{Bean1964} and exploiting magnetostatic-electrostatic analogues.\cite{Genenko2009,Genenko2011}

\section{THEORETICAL MODEL}
\label{mod} 

Let us first define the magnetostatic problem by considering a cylindrical superconductor/paramagnet 
heterostructure of bilayer geometry, {\it viz.}, an infinitely extended type-II superconducting tubular 
wire of radius $R$ and thickness $d$ on an inner, or outer, paramagnetic support of respective thickness 
$D$, buffered by an infinitesimally thin non-magnetic layer in between, and subject to a transverse 
magnetic field with strength $H_a$. We choose dimensions that second-generation coated conductors typically 
display,\cite{Claassen2008,Suenaga2008} {\it i.e.} $R=5 \rm\: mm$, $d=2\rm\: \mu m$, and 
$D=250 \rm\: \mu m$, understanding that the paramagnetic support is delineated by a finite permeability $\mu$.
Since $d\ll R$, we ignore spatial variations of the induced current on a length scale less than $d$ and,
for mathematical convenience, regard the superconducting tube as infinitesimally thin too, so that 
its physical state can be characterized by the sheet current $J$ 
alone.\cite{Genenko2010,Mawatari2008,Genenko2009,Genenko2011,Brandt1993} In conformity with Bean's model 
of the critical state duly adapted to the geometry of the tube for a polar orientation of the applied 
magnetic field,\cite{Mawatari2011,Brandt1993} magnetic 
flux penetrates from both equatorial sides of the tube into two cylindrical segments, of angle
$2\gamma $, where the sheet current $J$ equals the constant $J_c$; flux-free regions prevail in the polar 
segments of the tube, where the normal component of the magnetic field $H_n$ disappears.

As shown elsewhere,\cite{Genenko2009} the physical state of the tube in the corresponding electrostatic
problem then is characterized by the surface charge density $\sigma $ alone, invoking two central,
dielectric parts of the tube, of angle $2\gamma $, where the surface charge density $\sigma $ 
adopts the (field-independent) critical value $\sigma_c$, contiguous to polar, metallic parts of the 
tube at zero electrostatic potential $\varphi$; the support is represented by a dielectric 
with a finite permittivity $\varepsilon $. The heterostructure itself is subject to an electric field 
perpendicular to the applied magnetic field, with strength $E_a$, generated by finite potential values
$\varphi=\pm\varphi_a $ at opposite (left and right) sides of the quadratic computation frame well 
encompassing the domain covered by a cross section of the heterostructure.

Therefore, the following analogues between the magnetostatic problem and the
electrostatic problem hold:\cite{Genenko2009,Genenko2011} $J_c=\alpha \sigma_c$ with a free 
constant $\alpha $, relating the critical sheet current to the critical surface charge density,  $\mu/\mu_0=\varepsilon_0/\varepsilon$ with the vacuum permeability $\mu_0$ and 
the vacuum permittivity $\varepsilon_0 $, linking the permeability of the paramagnetic support 
to the permittivity of the dielectric support, and $H_a=\alpha \varepsilon_0 E_a$, relating the 
strength of the applied magnetic field to the strength of the applied electric field. It is thus 
clear that lines of the magnetic field and equipotentials of the electric field coincide. Hence, 
for an oscillating applied magnetic field with amplitude $H_a$, the penetration of magnetic flux 
and the consequential dissipation of energy, per cycle and unit length of the tube, $U_{ac}$ 
can be ascertained, resorting to the quasistatic 
approach.\cite{Genenko2010,Genenko2009,Genenko2011,Brandt1993}

Computations run as follows: given $\varepsilon $ and $\gamma $ for an arbitrarily chosen non-zero 
potential value $\varphi _a$, the critical value $\sigma_c$ is varied until a 
continuous profile of the surface charge density $\sigma$ over the circumference of the tube is reached; 
procedure here performed with COMSOL, a commercial finite-element software package not originally
designed for modelling superconducting states. On introducing the characteristic magnetic
field $H_c=J_c/\pi$, the half-angle of flux penetration $\gamma $ then is given in terms of the 
ratio $H_a/H_c=\pi\varepsilon_0 E_a/\sigma_c$, independent of the constant $\alpha$ and the potential 
value $\varphi_a$. The normalized hysteretic ac loss appears as\cite{Genenko2010,Genenko2009}
\begin{equation}
\label{loss}
U_{ac}/H_a^2 =  \left( 16\mu_0 \sigma_c  R^2 / \varepsilon_0 E_a^2 \right)\lim_{d\rightarrow 0}
\int_{0}^{\gamma} d\phi \int_{\phi}^{\gamma} d\phi'\, \bar{E}_t \left( \phi' \right),
\end{equation}
independent of $\alpha$ and $\varphi_a$ too, since the electric quantities $\sigma_c,E_a$, and
$\bar{E}_t$ in this equation, where $\bar{E}_t$ denotes the average tangential component of the 
electric field on the surface of the tube, all scale with the magnitude of $\varphi_a$.
We comment that the above analysis implies a neglect of the (small) contribution to the hysteretic 
ac loss arising from the tangential component of the magnetic field $H_t$, consistent with a threshold 
for the half-angle of flux penetration $\gamma$.

\begin{figure}[b]\center
\includegraphics[width=4.2cm]{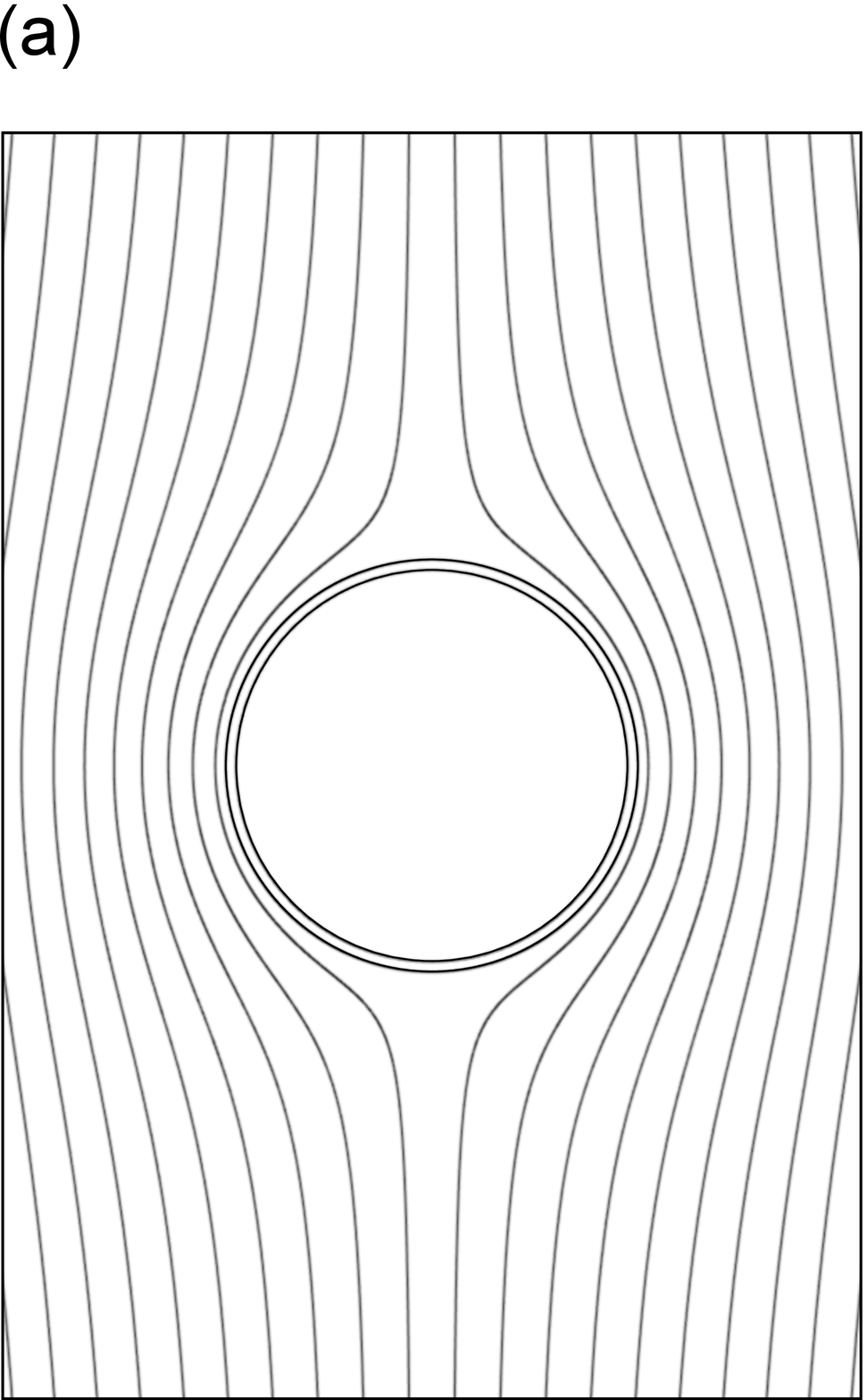}
    \hspace{1.0cm}
\includegraphics[width=4.2cm]{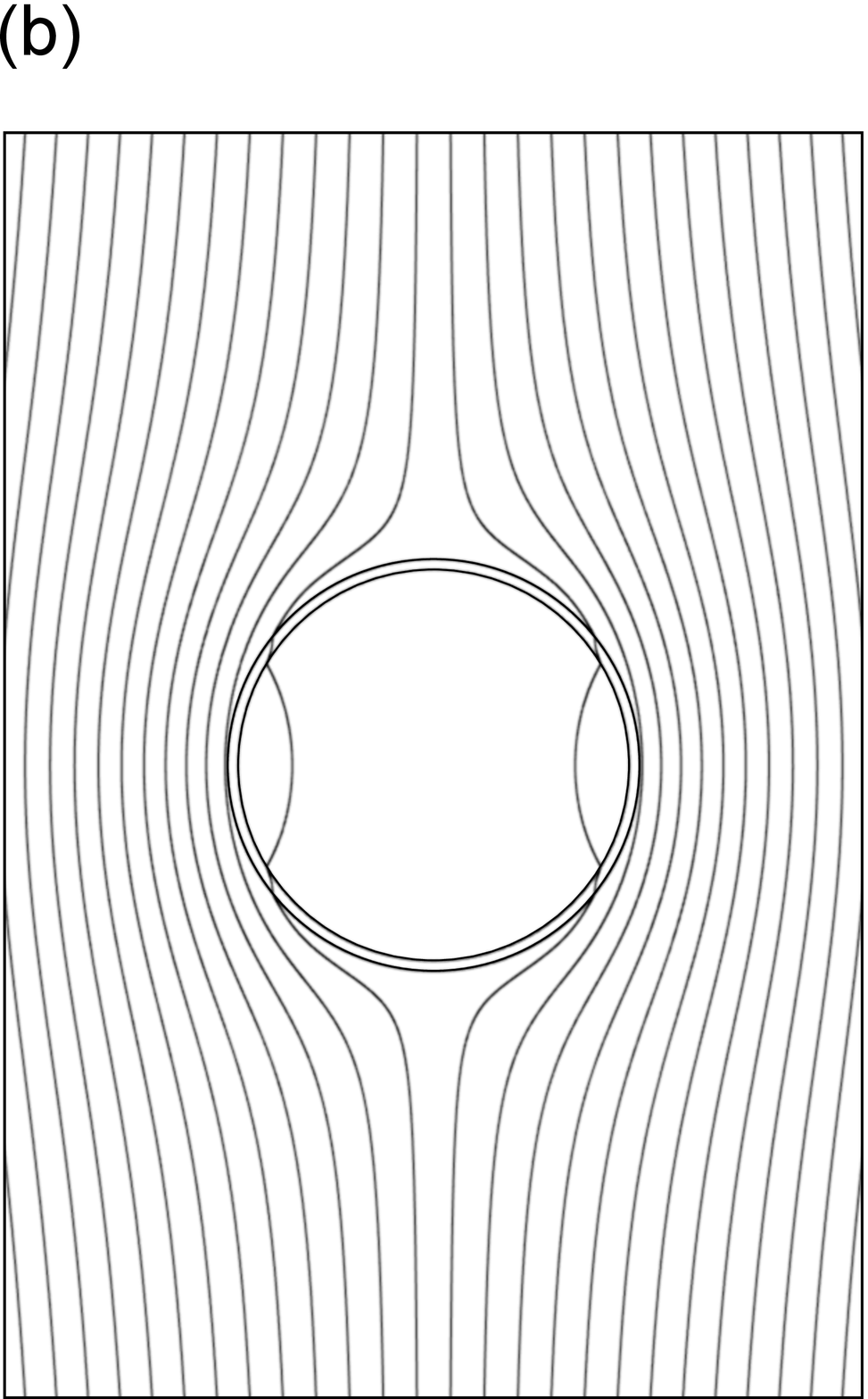}
 \hspace{0.0cm}
\includegraphics[width=4.2cm]{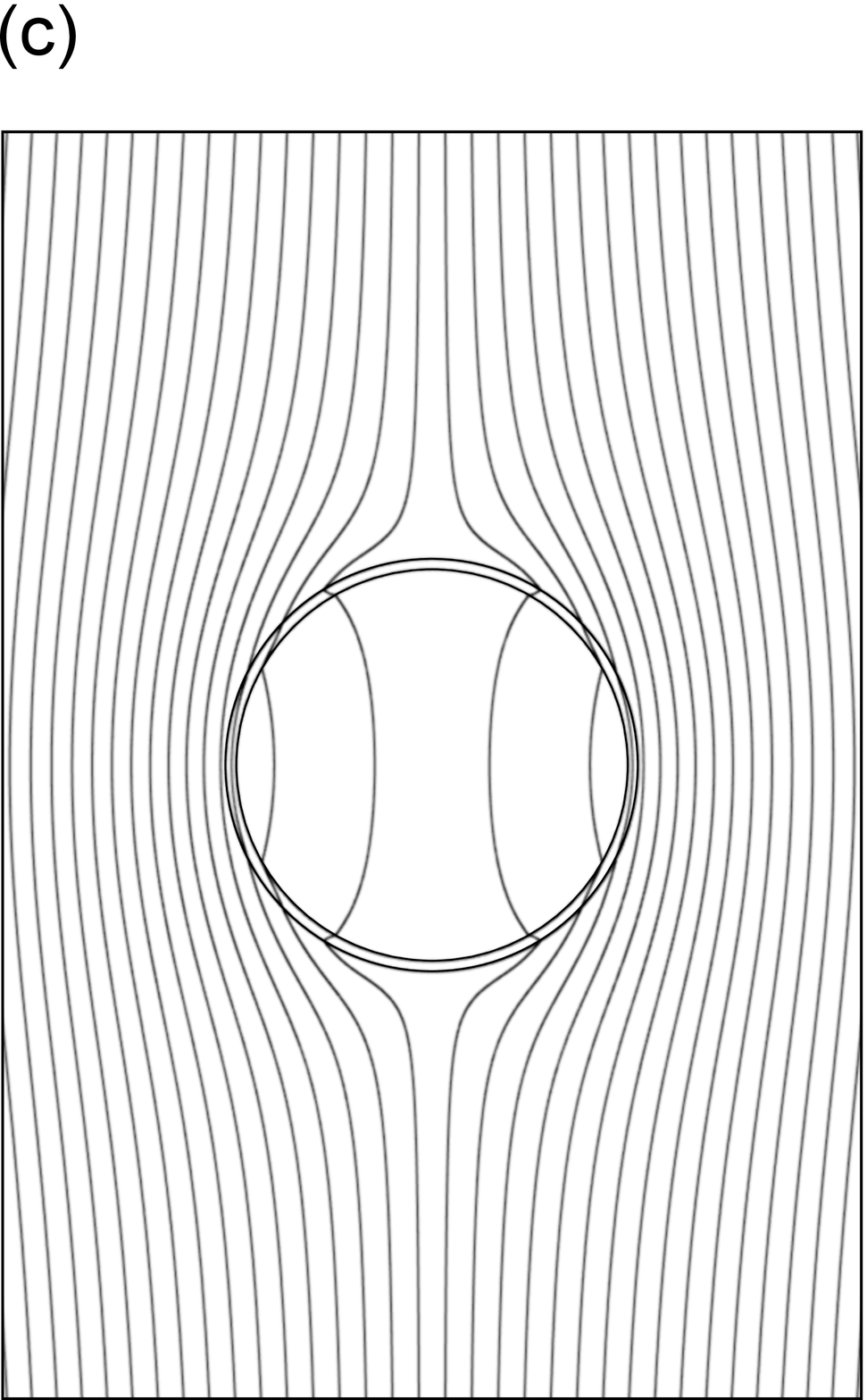}
\caption{\label{tub-inner} (Colour online) Lines of the magnetic field around the superconducting 
tubular wire coated with an {\it inner} paramagnetic support of relative permeability 
$\mu/\mu_0 = 10$, when the normalized amplitude of the magnetic field (a) $H_a/H_c = 1.5$, 
(b) $H_a/H_c = 2.0$, and (c) $H_a/H_c = 2.5$. The support together with the 
tube is indicated by black contour lines. }
\end{figure}

\section{NUMERICAL RESULTS}
\label{results} 

The following numerical results address supports 
placed on either side of the superconducting tube.

\subsection{Inner paramagnetic support  }
\label{in-sup} 

Referring to the case of an inner paramagnetic support, Fig.~\ref{tub-inner} illustrates 
the distribution of the magnetic field around the magnetically coated superconducting tubular 
\begin{figure}[b]\center
\includegraphics[width=8.0cm]{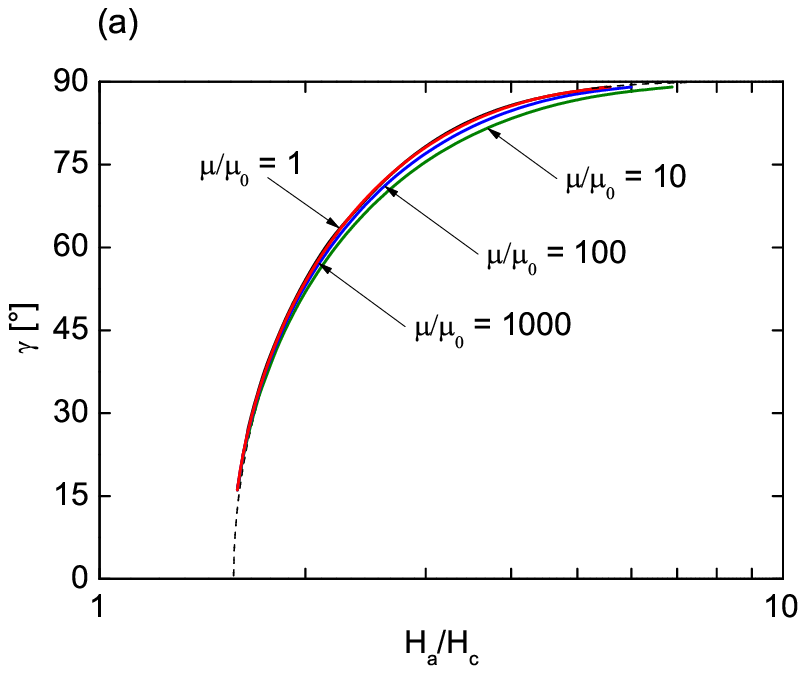}
 \hspace{0.0cm}
\includegraphics[width=8.0cm]{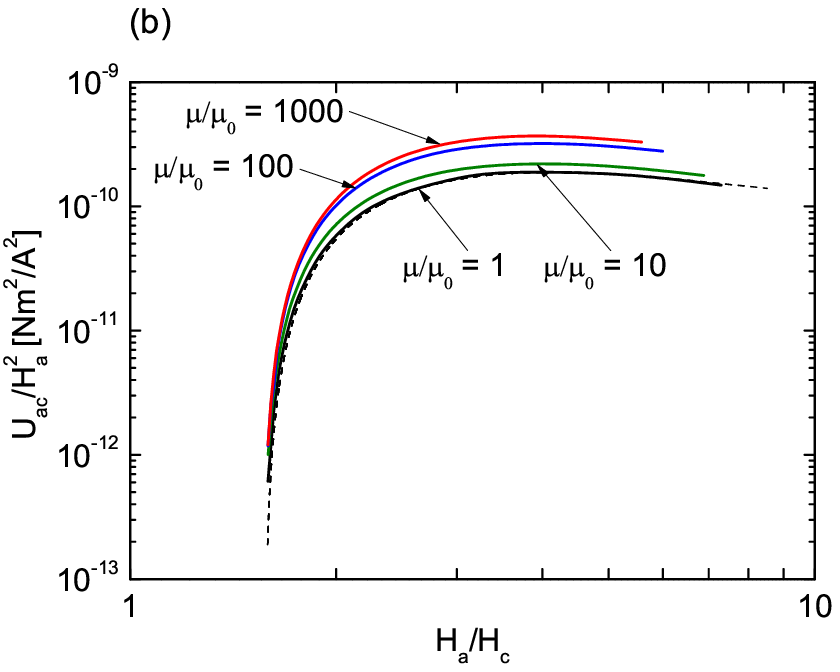}
\caption{\label{inner-data} (Colour online)
Dependence of (a) the half-angle of flux penetration $\gamma$ and (b) the normalized 
hysteretic ac loss $U_{ac}/H_a^2$ on $H_a/H_c$, the normalized amplitude of the magnetic 
field applied to the magnetically coated superconducting tubular wire, in the computationally 
accessible regimes, for four different values of the relative permeability of the {\it inner} 
paramagnetic support $\mu/\mu_0$ identified on the curves. The dashed lines represent 
analytical results for the isolated superconducting tube.\cite{Mawatari2011}}
\end{figure}
wire for a fixed permeability and three progressive values of the amplitude of the applied magnetic field. 
At $H_a/H_c = 1.5$, the interior of the wire is completely shielded from the 
applied magnetic field by the superconducting tube on top of the paramagnetic support: the lines of 
the magnetic field pass the wire outside (Fig.~\ref{tub-inner}(a)). As the amplitude of the 
applied magnetic field is increased to $H_a/H_c = 2.0$, magnetic flux starts to enter the 
superconductor constituent tangentially from both equatorial sides, threading two cylindrical 
segments of the tube and permeating the paramagnetic support, with refraction of the lines of 
the magnetic field at the interfaces therein, before the interior of the wire accommodates 
the flux (Fig.~\ref{tub-inner}(b)); an effect which, for $H_a/H_c = 2.5$, gets more extended 
still, demonstrating intensified refraction towards the poles (Fig.~\ref{tub-inner}(c)). 
Clearly, when penetration of magnetic flux occurs, the paramagnetic support is bound to exert 
an influence on the distribution of the magnetic field. The overall impact, however, remains 
weak due to the protecting action of the tube.

The variation of the half-angle of flux penetration with the amplitude of the applied
magnetic field, depicted in Fig.~\ref{inner-data}(a) for a range of values of the permeability, 
confirms these traits: penetration of magnetic flux sets in at $H_a/H_c=\pi/2$, like for an 
isolated superconducting tube;\cite{Mawatari2011} the half-angle of flux penetration rises 
monotonically with a tendency to saturation in a fully flux-filled state, as the amplitude 
of the applied magnetic field augments. The dependence on the permeability, by comparison, 
is non-monotonic and rather weak. 
The variation of the hysteretic ac loss with the amplitude 
of the applied magnetic field shown in Fig.~\ref{inner-data}(b) corroborates the threshold 
for the onset of this loss followed by a sharp monotonic rise, with a turn to a shallow 
maximum. 
A slight, but distinctly monotonic, increase occurs when the permeability is raised; 
even signs of saturation at higher values of the permeability exist, as predicted 
before.\cite{Genenko2000,GenenkoJAP2002}

\subsection{Outer paramagnetic support  }
\label{out-sup}

\begin{figure}[t]\center
\includegraphics[width=4.2cm]{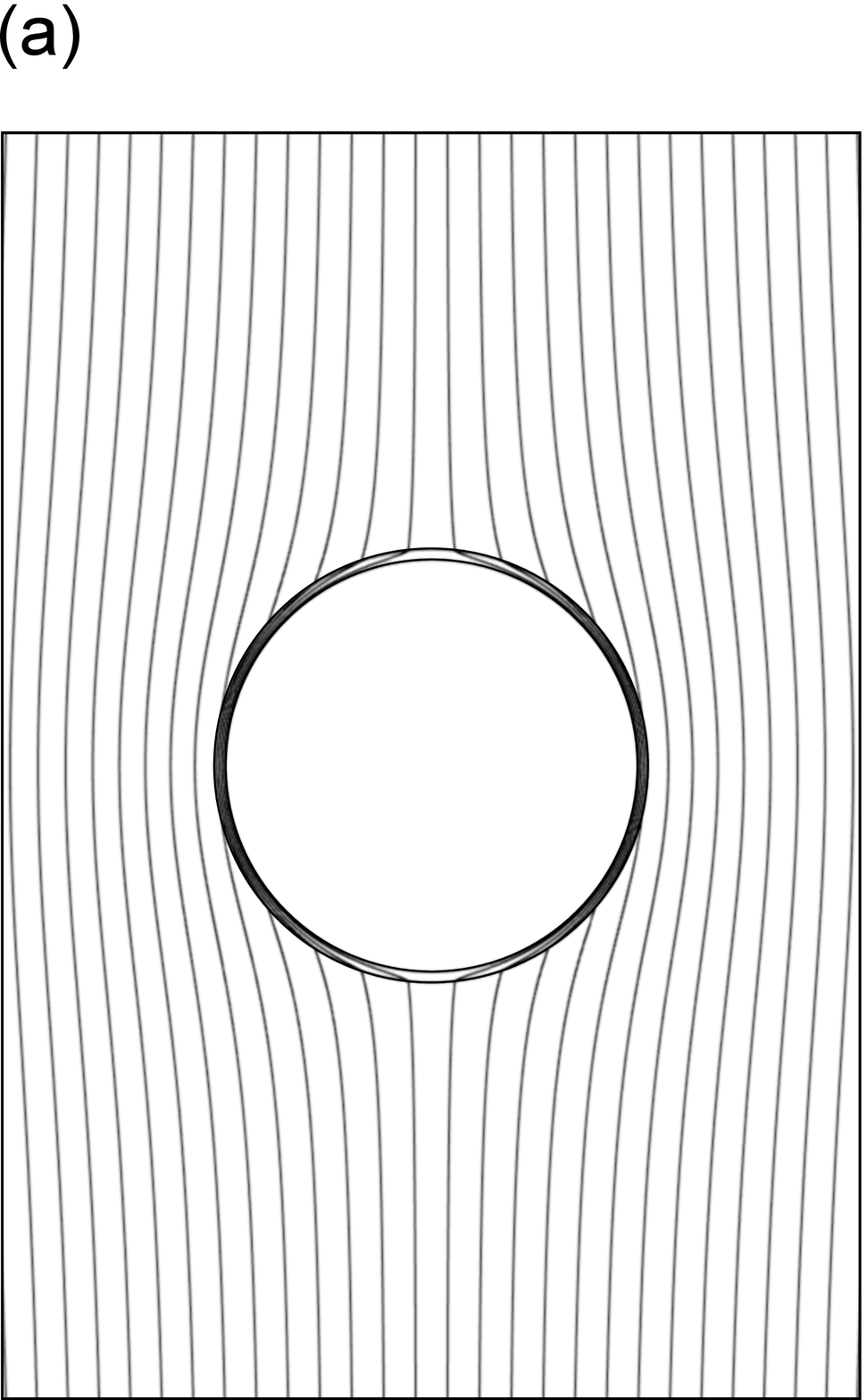}
    \hspace{1.0cm}
\includegraphics[width=4.2cm]{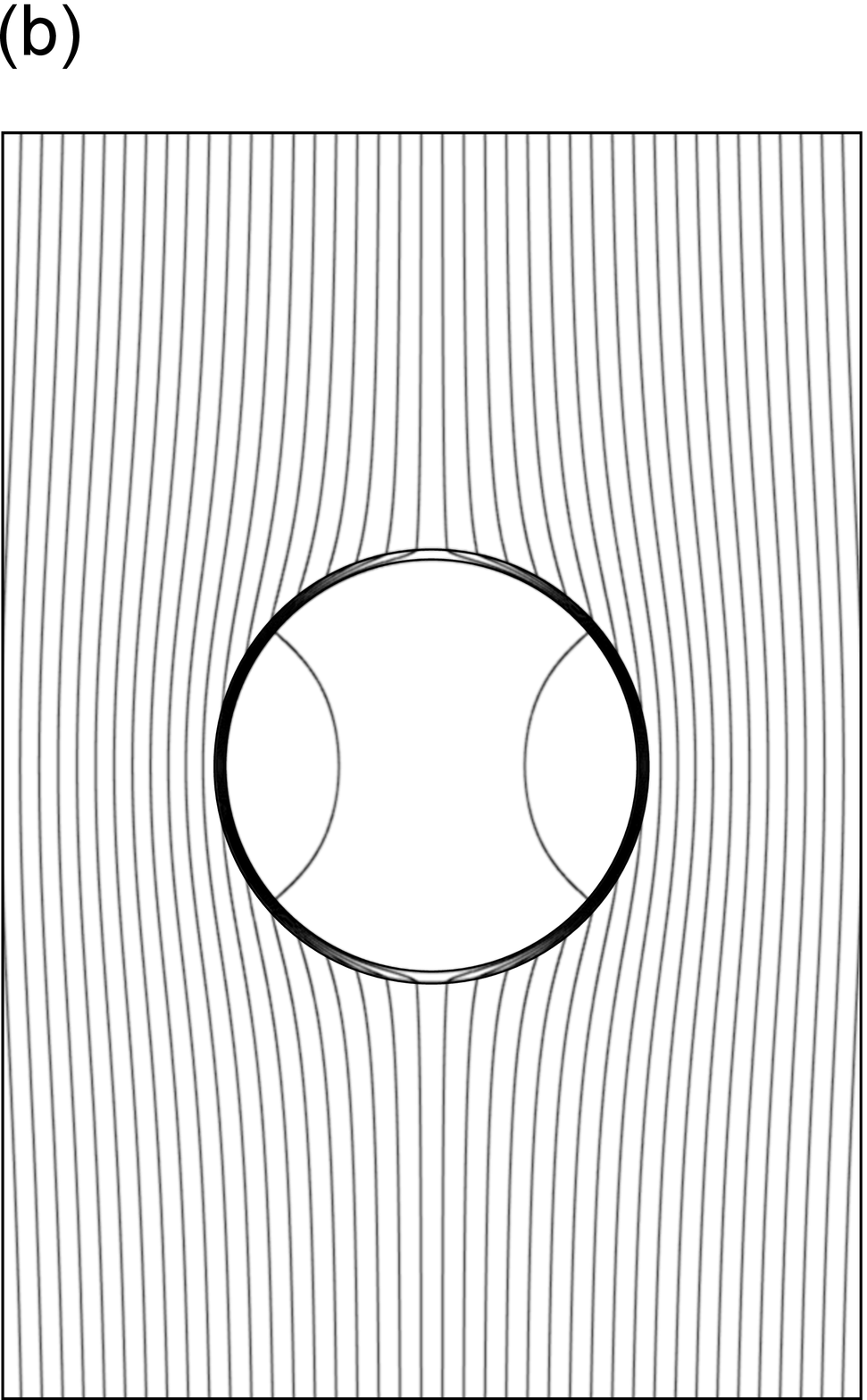}
 \hspace{0.0cm}
\includegraphics[width=4.2cm]{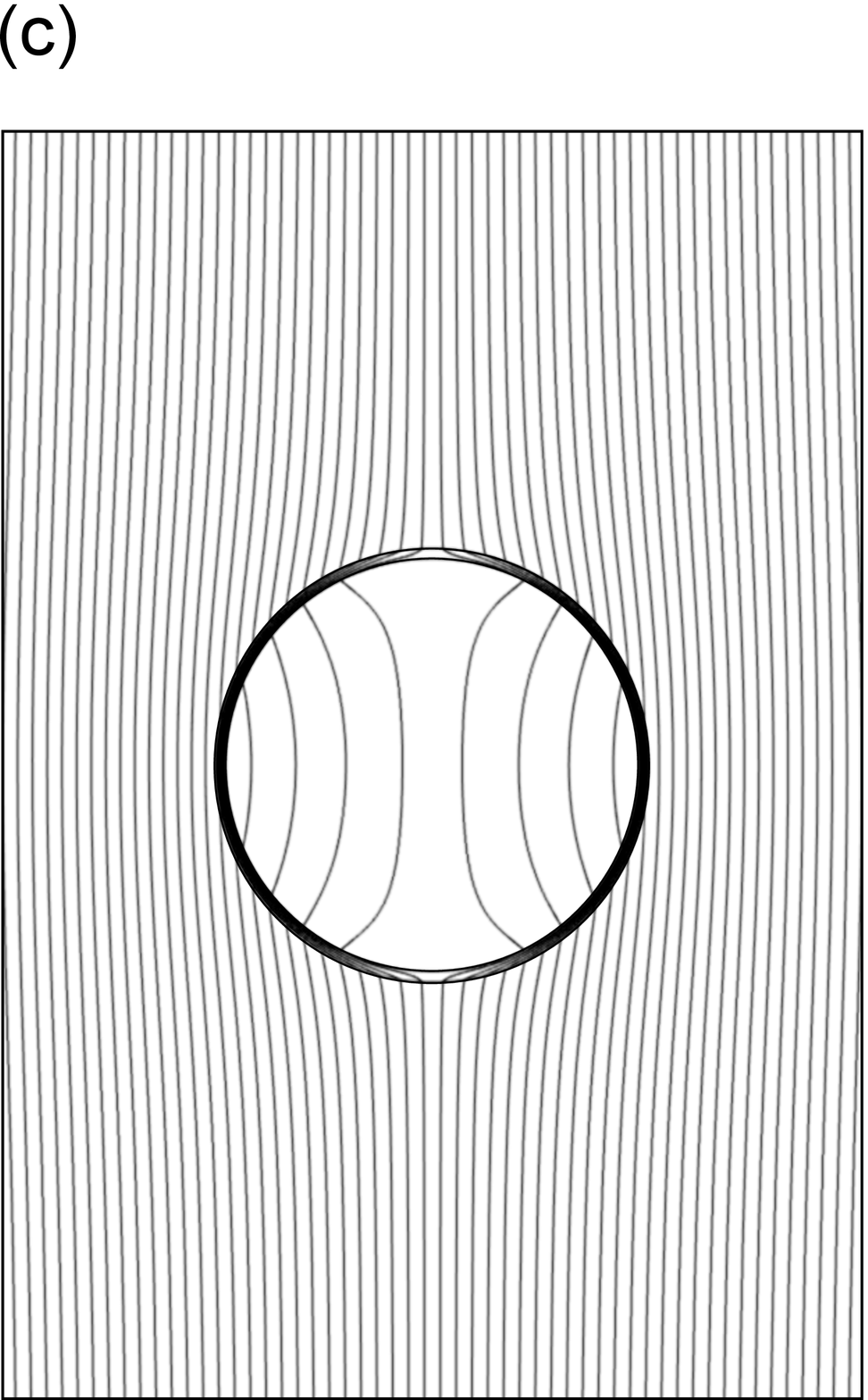}
\caption{\label{tub-outer}(Colour online) Lines of the magnetic field around the superconducting 
tubular wire coated with an {\it outer} paramagnetic support of relative permeability 
$\mu/\mu_0 = 10$, when the normalized amplitude of the magnetic field (a)  $H_a/H_c = 2.0$, 
(b) $H_a/H_c = 3.0$, and (c) $H_a/H_c = 4.0$. The support together with the tube is 
indicated by black contour lines.}
\end{figure}

Evidently, matters look different in the case of an outer paramagnetic support. This 
first concerns the distribution of the magnetic field around the magnetically coated
superconducting tubular wire portrayed in Fig.~\ref{tub-outer}, 
again for the permeability 
set and three progressive values of the amplitude of the applied magnetic field. 
At $H_a/H_c=2.0$, the interior of the wire is still completely shielded from the applied
magnetic 
field by the paramagnetic support on top of the superconducting tube: the lines of the 
magnetic field are refracted at the wire's outer surface and guided around inside the 
paramagnetic support (Fig.~\ref{tub-outer}(a)). As the amplitude of the applied magnetic 
field is increased to $H_a/H_c=3.0$, magnetic flux starts to enter the superconductor 
constituent tangentially from both equatorial sides, threading two cylindrical segments of 
the tube, with refraction of the lines of the magnetic field at the wire's inner surface
too, before the interior of the wire accommodates the flux (Fig.~\ref{tub-outer}(b)); an 
effect which, for $H_a/H_c=4.0$, gets more pronounced still, exhibiting intensified 
refraction towards the poles (Fig.~\ref{tub-outer}(c)). The paramagnetic support here 
always plays a prominent role in the distribution of the magnetic field. Its shielding 
capacity defines an effective, reduced field $H_{\mu}$ that acts on the wire's 
superconductor constituent.

\begin{figure}[t]\center
\includegraphics[width=8.18cm]{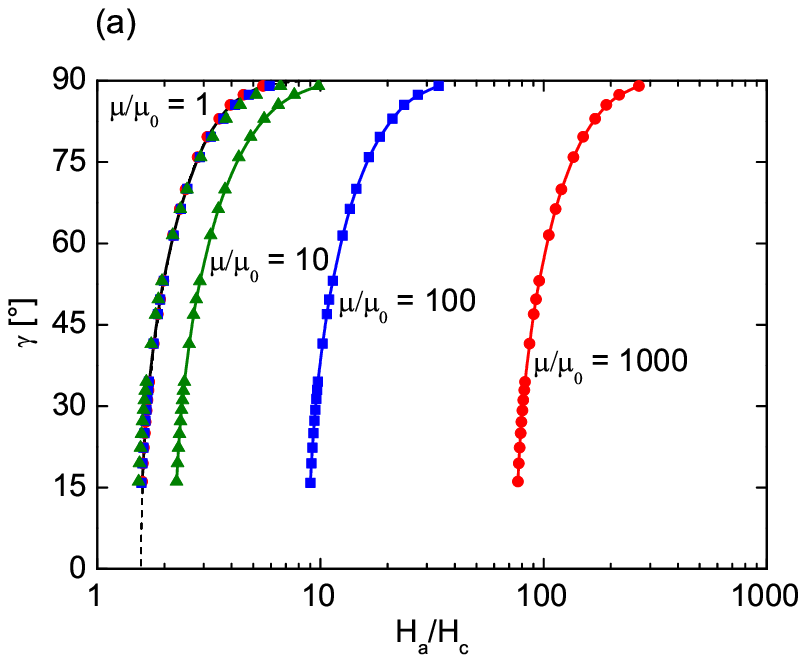}
 \hspace{0.0cm}
\includegraphics[width=8.18cm]{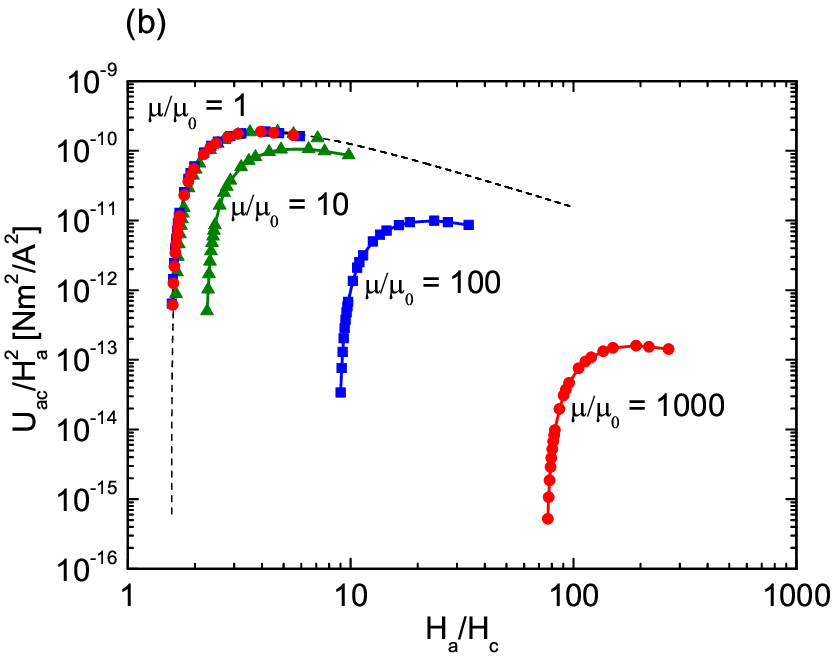}
\caption{\label{outer-data} (Colour online)
Dependence of (a) the half-angle of flux penetration $\gamma$ and (b) the normalized 
hysteretic ac loss $U_{ac}/H_a^2$ on $H_a/H_c$, the normalized amplitude of the magnetic 
field applied to the magnetically coated superconducting tubular wire, in the computationally 
accessible regimes, for four different values of the relative permeability of the {\it outer}
paramagnetic support $\mu/\mu_0$ identified on the curves. The full triangles, squares and
circles elucidate the scaling properties of the respective physical observables; the dashed 
lines represent analytical results for the isolated superconducting tube.\cite{Mawatari2011}}
\end{figure}

The dissimilarity of behaviour carries over to the variation of the half-angle of flux
penetration with the amplitude of the applied magnetic field, depicted in 
Fig.~\ref{outer-data}(a) for a range of values of the permeability: assuming the permeability 
of a vacuum, penetration of magnetic flux starts at $H_a/H_c=\pi/2$, again like for an 
isolated superconducting tube,\cite{Mawatari2011} with a monotonic rise and a tendency towards 
saturation in a fully flux-filled state, as the amplitude of the applied magnetic field augments. 
Increasing the permeability to account for a paramagnetic support produces a duplication of 
this course, yet rescaled to higher values of the amplitude of the applied magnetic field;
a fact which directly reflects the enhanced shielding capacity of the support. 
Corresponding traits recur in the variation of the hysteretic ac loss with the amplitude of the 
applied magnetic field shown in Fig.~\ref{outer-data}(b). The familiar threshold for the onset of 
the hysteretic ac loss appears, followed by a sharp monotonic rise with a turn to a shallow 
maximum, given the permeability of a vacuum environment. Increasing the permeability to model 
a paramagnetic support thus yields a shift of the hysteretic ac loss to higher values of the 
amplitude of the applied magnetic field, but also a considerable reduction in extent. 
It should be noted that these results can be calibrated to high accuracy against 
the expressions
\begin{equation}
\label{gamma}
\gamma = \gamma \left( H_{\mu}/H_c \right)\phantom{0}
\end{equation}
with
\begin{equation}
\label{gamma-fit}
H_a/H_{\mu} = 1 + 0.048 \left( \mu/\mu_0 -1 \right)
\end{equation}
and
\begin{equation}
\label{scaled-loss}
U_{ac}/H_a^2 = 8\pi \mu_0 R^2 \lambda_{\mu} f\left( H_{\mu}/H_c \right)
\end{equation}
with
\begin{equation}
\label{scale-parameter}
\log{\lambda_{\mu}} = -0.325 \log^2 \left( \mu/\mu_0 \right),
\end{equation}
the chosen values of the geometrical parameters of the wire implied, making recourse to the
functional dependences $\gamma \left( H_{a}/H_c \right)$ and $f\left( H_{a}/H_c \right)$
deduced for the isolated superconducting tube, either from analytical 
theory\cite{Mawatari2011} or numerical analysis.

\section{DISCUSSION AND CONCLUSION}
\label{conc} 

There are two sources bringing about modifications of the hysteretic ac loss in the
superconducting tube for a given paramagnetic support: changes of the amplitude of the
applied magnetic field and adjustments of the orientation of the local magnetic field which 
acts on the superconducting tube, a steep angle with the surface of the tube resulting in 
an enhancement of the loss. The case of an inner paramagnetic support clearly testifies to 
the latter fact. On the other hand, both sources favour a reduction of the hysteretic ac 
loss in the case of an outer paramagnetic support by depressing and guiding the magnetic flux, 
consistent with the findings of a previous analysis, which noted the factorization of the 
shielding effects of the superconducting and paramagnetic constituents before.\cite{Lousberg2010} 
The decrease of the hysteretic ac loss in this configuration certainly involves a more subtle 
interaction between the two types of constituents. Shielding of the transverse applied magnetic 
field by the paramagnetic support alone would generate a homogeneous magnetic field inside the 
tube, like for a massive cylindrical set, the inverse of its reduced strength $H_a/H_{\mu}$ 
unfolding a nonlinear dependence on the permeability $\mu$.\cite{Genenko2004} The magnetic 
field in the interior of the tubular wire configuration discussed here, however, proves to be 
inhomogeneous and strongly redistributed due to the presence of the superconductor constituent, 
the inverse of the effective, reduced field $H_a/H_{\mu}$ extracted from the numerical 
simulations demonstrating a perfectly linear dependence on $\mu$, as eqn.~(\ref{gamma-fit}) 
states. The variation of the shielding capacity of the outer paramagnetic support with $\mu$ 
in turn admits tuning the hysteretic ac loss by a judicious choice of $\mu$.

Attention should be paid to the circumstance that, for a transverse applied magnetic field,
the shielding effect of an outer paramagnetic support in the superconducting tubular wire
configuration is usually strong and strictly advantageous regarding the hysteretic ac loss,
unlike that of a paramagnetic support in a planar bilayer superconductor/paramagnet
heterostructure which is comparatively weak and either beneficial or
detrimental, depending on the geometrical and material characteristics of the magnetic
environment and the amplitude of the applied magnetic field.\cite{Genenko2011,GomorySUST2010} 
The reason for this significant difference is of geometrical sort. The component of the 
magnetic field penetrating into the flat bilayer heterostructure is oriented normal to the 
plane of the structure; due to the paramagnetic support guiding the magnetic flux, it may 
either be enhanced or reduced. By contrast, the component of the magnetic field 
penetrating into the superconducting tube at its equatorial line is oriented tangential to 
the surface of the tube; an outer paramagnetic support guiding the magnetic flux can only 
promote this grazing orientation, thus reducing the field-effect on the superconductor 
constituent.

In conclusion, our finite-element simulations reveal effects of a paramagnetic support on
hysteretic ac losses in a tubular superconductor/paramagnet heterostructure subject to an
oscillating transverse magnetic field. Accordingly, one-sided magnetic shielding of the
superconducting tube by a coaxial paramagnetic support gives rise to a slight increase of
hysteretic ac losses as compared to those for a vacuum environment, when the support is
placed inside; a spectacular shielding effect with a possible reduction of hysteretic ac 
losses by orders of magnitude, however, ensues, depending on the magnetic permeability 
and the amplitude of the applied magnetic field, when the support is placed outside.

\bibliography{aipsamp}

\end{document}